\documentclass[preprint2]{aastex}

\shorttitle{peculiar SNe Ia from SD model} \shortauthors{Meng \&
Han}

\begin{document}


\title{Why Are Peculiar Type Ia Supernovae More Likely to Show
the Signature of a Single-degenerate Model?}


\author{Xiang-Cun Meng$^{\rm 1,2,3}$, Zhan-Wen Han$^{\rm 1,2,3}$}
\affil{$^{\rm 1}$Yunnan Observatories, Chinese Academy of Sciences, 650216 Kunming, PR China\\
$^{\rm 2}$ Key Laboratory for the Structure and Evolution of
Celestial Objects, Chinese Academy of Sciences, 650216 Kunming, PR
China\\
$^{3}$Center for Astronomical Mega-Science, Chinese Academy of
Sciences, 20A Datun Road, Chaoyang District, Beijing, 100012, P.
R. China} \email{xiangcunmeng@ynao.ac.cn}





\begin{abstract}
Although type Ia supernovae (SNe Ia) are very useful in many
astrophysical fields, their exact progenitor nature is still
unclear. A basic method to distinguish the different progenitor
models is to search the signal from the single-degenerate (SD)
model, e.g., the signal for the existence of a nondegenerate
companion before or after supernova explosion. Observationally,
some SNe Ia show such signals, while the others do not. Here, we
propose a universal model to explain these observations based on
the spin-up/spin-down model, in which a white dwarf (WD) will
experience a spin-down phase before supernova explosion, and the
spin-down timescale is determined by its initial mass, i.e., the
more massive the initial WD, the shorter the spin-down timescale
and then the more likely the SN Ia is to show the SD signature.
Therefore, our model predicts that the SNe Ia from hybrid
carbon-oxygen-neon WDs are more likely to show the SD signature
observationally, as some peculiar SNe Ia showed.
\end{abstract}





\section{Introduction}\label{sect:1}
Although Type Ia supernovae (SNe Ia) show their importance in many
astrophysical fields, e.g. as standard candles to measure
cosmological parameters (\citealt{RIE98}; \citealt{PER99}), a
decades-long debate is endless on their progenitors
(\citealt{HN00}; \citealt{WANGB12}). A consensus has been achieved
that the thermonuclear explosion of a carbon-oxygen white dwarf
(CO WD) in a binary system produces an SN Ia (\citealt{HF60}).
Based on the companion nature of the mass accreting WDs, the
progenitor models of SNe Ia were divided into two basic scenarios:
one is the single-degenerate (SD) model where the companion is a
normal star, i.e. a main-sequence or a slightly evolved star
(WD+MS), a red giant star (WD+RG) or a helium  star (WD + He
star), the other involving the merger of two CO WDs is the double
degenerate (DD) model (\citealt{WANGB12}; \citealt{MAOZ14}).

A basic method to distinguish the different models is to search
the signature from the nondegenerate companion before or after
supernova explosion, e.g. to search the surviving companion in a
supernova remnant, to detect the UV excess from the interaction
between supernova ejecta and the companion or to detect the
progenitor system directly from the archival images before
supernova explosion (\citealt{WANGB12}; \citealt{MAOZ14}). Many
efforts are performed to search such signals, in which some
clearly show the signals (\citealt{FOLEY14}; \citealt{MCCULLY14};
\citealt{CAOY15}), while the others do not (\citealt{LIWD11};
\citealt{GONZALEZ12}; \citealt{KEERZENDORF14};
\citealt{OLLING15}). The simplest explanation is that some SNe Ia
originate from the SD systems, and the others are from the DD
ones. However, an interesting puzzle is that the SNe Ia exhibiting
the companion signal tend to be the subluminous objects with low
ejecta velocities, e.g. SN 2008ha, SN 2012Z, and iPTF14atg
(\citealt{FOLEY14}; \citealt{MCCULLY14}; \citealt{CAOY15}) which
are classified as peculiar SNe Ia (SN 2002cx-like or SN
2002es-like objects, \citealt{LIWD03}; \citealt{GANESHALINGAM12}),
although some normal SNe Ia also show the UV excess from the
collision between supernova ejecta and the companion, e.g. SN
2012cg and 2017cbv (\citealt{MARION16}; \citealt{HOSSEINZADEH17}).
On the contrary, the SNe Ia without companion signals tend to be
normal SNe Ia, e.g SN 1006, SN 2011fe, KSN 2012a, and KSN 2011b
(\citealt{LIWD11}; \citealt{OLLING15}; \citealt{KATSUDA17}).
However, it must be emphasized that SN 2012cg and 2017cbv have a
large binary separation at the moment of supernova explosion
(\citealt{MARION16}; \citealt{HOSSEINZADEH17}), and then are very
likely to have relatively massive initial CO WDs since the more
massive the initial WD, the more likely to explode in a large
binary separation for a SN Ia according to detailed binary
evolution calculations (e.g. Figure 12 in \citealt{MENGXC17a}).

Another strong piece of evidence favoring the SD model is the
detection of circumstellar material (CSM) in the spectrum of SNe
Ia (\citealt{HAMUY03}; \citealt{PATAT07}; \citealt{STERNBERG11};
\citealt{DILDAY12}). The SNe Ia showing the strong CSM signal are
classified as SNe Ia-CSM (\citealt{SILVERMAN13}). Both SN Ia-CSM
and SN 2002cx-like objects present the spectra similar to SN 1991T
and originate from young populations (\citealt{SILVERMAN13};
\citealt{FOLEY13}). The overluminous 1991T-like events also favor
SD systems with significant mass loss before supernova explosion,
even as high as $\sim10^{\rm -5}~{\rm M_{\odot}/yr}$
(\citealt{FISHER15}; \citealt{KATSUDA15}), where the high
mass-loss rate likely indicates a massive initial WD (e.g. Fig.4
in \citealt{MENGXC17a}). Considering that SN 2002cx-like and SN
Ia-CSM objects share some common properties and the explosions of
hybrid carbon-oxygen-neon (CONe) WDs appear rather heterogeneous,
\citet{MENGXC17b} suggested that both subclasses could originate
from the SD systems with hybrid CONe WDs. Although it cannot be
completely excluded that the both subtypes have different origins,
their model may reproduce the number ratio of SN Ia-CSM to SN
2002cx-like objects and the total contribution of the peculiar SNe
to all SNe Ia. This suggestion is based on a new-version SD model,
i.e. the common envelope wind (CEW) model (\citealt{MENGXC17a}),
where the CE mass distribution at the moment of supernova
explosion is double peaked. The SNe Ia-CSM correspond to those
with massive CE, while 2002cx-like SNe are from those exploding in
less massive or no CE. \citet{MENGXC17b} suggest that different
explosion environment is the main reason why 2002cx-like and SN
Ia-CSM objects seem quite different. In addition, the different
cooling times of the WDs before accretion occurring for two
subclasses could also play a key role in their different
properties (see the discussions in \citealt{MENGXC17b}). In
particular, the double peak CE mass distribution provides a
potential explanation for the fact that no transitional event
between SN 2002cx-like and SN Ia-CSM objects is discovered.
Although arguments exist on whether or not a hybrid CONe WD may
form and carbon ignition may occur in the hybrid WD, the chemical
evolution of dwarf spheroidal galaxies may even provide some
indirect support for their explosion (\citealt{LECOANET16};
\citealt{BROOKS17}; \citealt{CESCUTTI17}).

Compared with CO WDs, CONe WDs are relatively massive
(\citealt{CHENMC14}), as required by the normal SNe Ia with the SD
signal, e.g. SN 2012cg and 2017cbv. However, the environment
around the normal SNe Ia tends to be clear, e.g. SN 2011fe and SN
2014J, and then these SNe Ia are proposed to be from the DD
systems (\citealt{PATAT13}; \citealt{PEREZ14}). According to the
above discussion, an interesting question arises, i.e. why do the
SNe Ia with the SD signals tend to have massive initial WDs, while
those proposed to be from the DD systems tend to be normal SNe Ia?
May such a question be answered under a universal frame? Here, we
investigate these questions and show that a universal frame is
possible in principle.

In section \ref{sect:2}, we describe our method, and present the
calculation results in section \ref{sect:3}. We show discussions
and our main conclusions in section \ref{sect:4}.

\section{Method}
\label{sect:2}
To explain why no surviving companion is found in supernova
remnants, e.g. SNR 0509-67.5 (\citealt{SCHAEFER12}), the
spin-up/spin-down model is proposed, in which the WD is spun up by
accretion, and must experience a spin-down phase before it
explodes as an SN Ia (\citealt{DISTEFANO11}; \citealt{JUSTHAM11}).
Then, the SNe Ia may not reveal the SD signature for a long
spin-down timescale, which hints at a universal explanation of the
above observational facts. However, there are many uncertainties
on the model theoretically: (1) What is the fraction of the
angular momentum of the accreted material to transfer to the
accreting WD? (2) What is the mechanism to lose the angular
momentum for a nonaccreting WD? (3) How long is the spin-down
timescale, and (4) How is the unaccreted material ejected from the
system? Since the method here is just based on the conservation of
angular momentum, these uncertainties cannot significantly affect
our basic conclusion.

The spin-down timescale denotes a delay time of a rapidly rotating
WD from an initial fast rotation down to a critical angular
velocity, and its exact value is quite uncertain
(\citealt{DISTEFANO12}; \citealt{MENGXC13}). However, whether a
property from the SD model is observed or not heavily depends on
the spin-down timescale. Since the timescale is determined by the
initial angular velocity of the WD at the onset of the spin-down
phase, we may use the initial angular velocity to represent the
timescale. The angular velocity is dependent on the accretion
history of the WD. The angular momentum of the material accreted
along the equator of a WD is determined by
 \begin{equation}
{\rm d}J={\rm d}m R_{\rm WD}^{\rm 2}\omega_{\rm K},\label{dj}
  \end{equation}
where $R_{\rm WD}$ is the radius of the WD and $\omega_{\rm K}$ is
the Keplerian angular frequency at the surface of the WD:
 \begin{equation}
\omega_{\rm K}=\left(\frac{GM_{\rm WD}}{R_{\rm WD}^{\rm
3}}\right)^{\frac{1}{2}}.
  \end{equation}
The WD radius is determined by
 \begin{equation}
R_{\rm WD}=0.0115\sqrt{\left(\frac{M_{\rm Ch}}{M_{\rm
WD}}\right)^{2/3}-\left(\frac{M_{\rm WD}}{M_{\rm
Ch}}\right)^{2/3}}~R_{\odot},
  \end{equation}
as in \citet{TOUT97}, where $M_{\rm Ch}=1.44~M_{\odot}$ is the
Chandrasekhar mass. We assume that most of SNe Ia explode at the
same mass $M_{\rm WD}^{\rm SN}=1.4~M_{\odot}$, which is upheld by
both observations and theory (\citealt{SCALZO14};
\citealt{WANGB14}). However, the angular momentum obtained by the
accreting WD is probably much lower than that by Eq.~\ref{dj},
e.g. taking away via nova explosion, or accreting by a CE rather
than by a Keplerian disk (e.g. \citealt{MENGXC17a}). Then, the
total angular momentum obtained by the accreting WD may be
expressed by
 \begin{equation}
\Delta J=\int_{M_{\rm WD}^{\rm i}}^{M_{\rm WD}^{\rm SN}}\alpha
{\rm d}J=\int_{M_{\rm WD}^{\rm i}}^{M_{\rm WD}^{\rm SN}}\alpha
R_{\rm WD}^{\rm 2}\omega_{\rm K}{\rm d}m,
  \end{equation}
where $\alpha$ is a parameter of much lower than 1, which
indicates that the accreted material carries enough angular
momentum to spin-up the WD. To maintain the final angular velocity
to be smaller than Keplerian angular velocity, we take
$\alpha=0.003$ rather arbitrarily. Then, the final angular
momentum of the WD after the accretion phase is
 \begin{equation}
J_{\rm f}=I_{\rm WD}^{\rm SN}\omega_{\rm f}=\Delta J,
  \end{equation}
where $I_{\rm WD}^{\rm SN}$ is the moment of inertia at $M_{\rm
WD}^{\rm SN}=1.4~M_{\odot}$, and $\omega_{\rm f}$ is angular
frequency. We assume that $I_{\rm WD}^{\rm SN}$ is not dependent
on the $\omega_{\rm f}$, i.e. it is a constant for all rapidly
rotating WDs of $M_{\rm WD}^{\rm SN}=1.4~M_{\odot}$. The moment of
inertia is parameterized with a structural constant $\beta$
 \begin{equation}
I_{\rm WD}^{\rm SN}=\frac{2}{5}\beta M_{\rm WD}^{\rm SN}R_{\rm
WD}^{\rm 2}.
  \end{equation}
Here, we take $\beta=0.27$ (\citealt{ILKOV12}). Then,
 \begin{equation}
\omega_{\rm f}=\frac{\Delta J}{I_{\rm WD}^{\rm SN}}.
  \end{equation}
We take $\omega_{\rm f}$ as the initial value of angular velocity
at the onset of the spin-down phase. At the spin-down phase, a
rigid rotating WD may lose its rotational kinetic energy by
magnetodipole radiation or gravitational wave radiation, or even
magnetic braking. Here, we use $\omega_{\rm f}^{\rm 2}$ to
represent the rotational kinetic energy since $I_{\rm WD}^{\rm
SN}$ is assumed to be a constant.




\begin{figure}
\centerline{\includegraphics[angle=270,scale=.35]{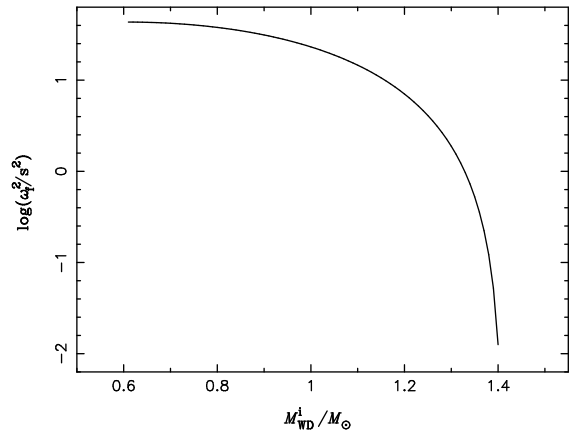}}
\caption{Rotational kinetic energy of a WD at the onset of the
spin-down phase as a function of its initial mass.}\label{spin}
\end{figure}

\section{Result}\label{sect:3}
\subsection{The Rotational Kinetic Energy}\label{sect:3.1}
In Fig.~\ref{spin}, we show the rotational kinetic energy of a WD
with its initial mass at the onset of the spin-down phase. As
expected, the rotational kinetic energy is heavily dependent on
the initial WD's mass, i.e. it quickly decreases with the initial
WD's mass. The rotational kinetic energy of a WD with $M_{\rm
WD}^{\rm i}=0.8~M_{\odot}$ is 20 times higher than that of a WD
with $M_{\rm WD}^{\rm i}=1.3~M_{\odot}$, where $0.8~M_{\odot}$ is
the most probable value and $1.3~M_{\odot}$ is the maximum mass
for a WD to produce an SN Ia (\citealt{CHENMC14}). At present, the
exact mechanism losing the rotational kinetic energy is still
unclear. For magnetodipole radiation, our result implies that the
spin-down timescale for a WD with $M_{\rm WD}^{\rm
i}=0.8~M_{\odot}$ may be more than 100 times as long as that of a
WD with $M_{\rm WD}^{\rm i}=1.3~M_{\odot}$, where the exact times
depend on the critical angular velocity triggering the
thermonuclear explosion (\citealt{ILKOV12}).

If the spin-down timescale is long enough, the SD signal will be
erased completely; otherwise, the SD signal will be expected. The
spin-down timescale is determined by the initial angular velocity
of a WD at the onset of the spin-down phase, and then by the
initial WD mass as shown in Fig.~\ref{spin}. In other words, the
spin-down timescale for a massive initial WD is shorter than a
less massive initial WD. Therefore, the more massive the initial
WD, the more likely the SN Ia is to show the \textbf{SD} signal.

\begin{figure}
\centerline{\includegraphics[angle=270,scale=.35]{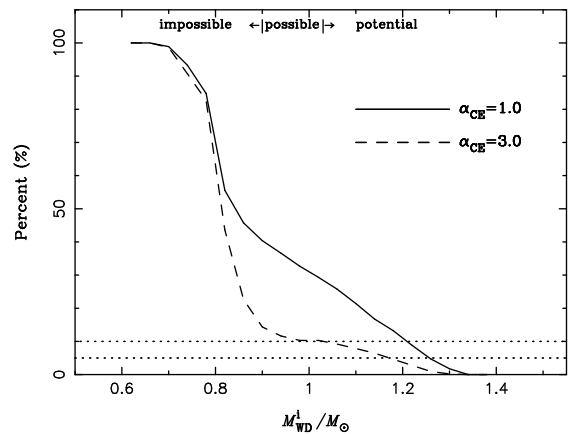}}
\caption{Proportion of SNe Ia with the SD signal as a function of
the initial WD mass for different common envelope ejection
efficiency, where the SNe Ia are from the CO and hybrid CONe WDs
(\citealt{MENGXC17a,MENGXC17b}). The horizontal lines correspond
to the values of 5\% and 10\%, respectively. The top short
vertical bars denote the WD mass regions that potentially,
possibly or impossibly show the SD signal.}\label{mwd}
\end{figure}
\subsection{The Proportion of SNe Ia with the SD Signal}\label{sect:3.2}
As discussed above, the more massive the initial WD, the more
likely the SN Ia is to show the SD signal. However, the proportion
of SNe Ia that have a potential to show the signarure would
heavily depend on a threshold value of the spin-down timescale.
Unfortunately, the threshold value for an SN Ia is completely
unclear. Since different initial WD masses correspond to different
spin-down timescales, we may use the different initial WD masses
to represent the different threshold values, and then study how
the proportion of SNe Ia with the SD signal relies on the
threshold value. Following the model grids in
\citet{MENGXC17a,MENGXC17b}, we performed two binary population
synthesis (BPS) calculations, where the method for the BPS
calculations is similar to that in \citet{MENGXC17a,MENGXC17b}.
Here, the WDs for SNe Ia include CO and hybrid CONe WDs. In
Fig.~\ref{mwd}, we show the proportion as a function of the
initial WD mass. As expected, the percentage decreases with
$M_{\rm WD}^{\rm i}$, i.e. the shorter the threshold value for the
spin-down timescale, the smaller the proportion of SNe Ia with the
SD signal is. In addition, the percentage sharply decreases around
$M_{\rm WD}^{\rm i}\simeq0.8~M_{\odot}$, which means that the
distribution of the WDs for SNe Ia peaks at $M_{\rm WD}^{\rm
i}\simeq0.8~M_{\odot}$. This implies that the threshold value of
the spin-down timescale would be shorter than that represented by
$M_{\rm WD}^{\rm i}=0.9~M_{\odot}$, since most SNe Ia do not
present the SD signal.

However, it should be emphasized that not all SNe Ia from massive
initial WDs must show the SD signal. For example, some SNe Ia from
massive initial WDs may have a very clear environment, or a less
massive companion at the moment of supernova explosion
(\citealt{MENGXC17a}). So, it will be very difficult to detect the
CSM around such SNe Ia, or to detect the UV excess from the
interaction between supernova ejecta and the companion. Moreover,
the UV excess is highly view-angel-dependent, which may
significantly reduce the possibility further (\citealt{KASEN10}).
Therefore, the proportion shown in Fig.~\ref{mwd} is just a
conservative upper limit.

As discussed in section \ref{sect:1}, the peculiar SNe Ia are more
likely to show the SD signal, but the contribution of the peculiar
SNe Ia to all SNe Ia is still uncertain (\citealt{LIWD11b};
\citealt{FOLEY13}). However, it is very possibly between 5\% and
10\% (\citealt{MENGXC17b}). In Fig.~\ref{mwd}, we also plot two
horizontal lines to show 5\% and 10\%. Corresponding to the value
of 10\%, the initial WDs must be more massive than $\sim1.03$ or
$\sim1.21~M_{\odot}$ relying on the CE ejection efficiency
($\alpha_{\rm CE}$). $1.03~M_{\odot}$ is close to the upper
boundary for CO WDs (\citealt{CHENMC14}). Compared with the SNe Ia
from the CO WDs, those from the hybrid CONe WDs are more likely to
show the SD signal for their higher initial mass. The SNe Ia with
CONe WDs are proposed to present the properties of peculiar SNe Ia
(\citealt{MENGXC14,MENGXC17b}; \citealt{KROMER15}). Such a
proposal obtains a further support from the fact that peculiar SNe
Ia have a higher probability of showing the SD signals.

In addition, our results do not exclude SNe Ia from massive CO WD
to show the SD signal since the exact proportion of SNe Ia with
the SD signal is unclear. If 15\% of SNe Ia have a potential to
show the SD signal, the threshold value of the initial WD mass is
$\sim0.9~M_{\odot}$ ($\alpha_{\rm CE}=3.0$) or $1.15~M_{\odot}$
($\alpha_{\rm CE}=1.0$), which is a value larger than that of most
of SNe Ia. This may explain why some normal SNe Ia exhibit the UV
excess from the interaction between supernova ejecta and the
companion, e.g. it is very possible that normal SN 2012cg and
2017cbv will have massive initial WDs (\citealt{MARION16};
\citealt{HOSSEINZADEH17})

\section{Discussions and Conclusions}\label{sect:4}
In this paper, we propose that whether an SN Ia will reveal the SD
signature is determined by its initial WD mass, based on the
spin-up/spin-down model. A massive WD only needs to accrete a
small amount of the material to reach the Chandrasekhar limit, and
then obtains a small amount of angular momentum. Such a WD will
take a shorter spin-down timescale for a slow rotation to get the
condition triggering an SNe Ia. Thus, the more massive the initial
WD, the more likely the SD signal is to be observed. As shown in
Fig.~\ref{mwd}, the SNe Ia from hybrid CONe WDs are more likely to
show the SD signal. Such SNe Ia are proposed to present the
properties of the SN 2002cx-like and SN Ia-CSM events
(\citealt{MENGXC14}; \citealt{KROMER15}; \citealt{MENGXC17b}).
This may explain why peculiar objects are more likely to show the
SD signal. Even though the SNe Ia are normal, it is still more
possible to show the SD signal for those with massive initial WDs,
e.g. SN 2012cg and 2017cbv (\citealt{MARION16};
\citealt{HOSSEINZADEH17}).

Observationally, most SNe Ia do not show the SD signal, which may
also be explained by our suggestion. In Fig.~\ref{mwd}, most SNe
Ia have an initial CO WD with $< 0.9~M_{\odot}$. If their
spin-down timescale is long enough, the properties predicted by
the SD model will be erased, and then most of the SNe Ia will not
show the SD signal. Our results imply that the threshold value of
the spin-down value for the SD signal corresponds to the one with
$M_{\rm WD}^{\rm i}\geq0.9~M_{\odot}$. Based on a semi-empirical
method, \citet{MENGXC13} found that the spin-down timescale is
shorter than a few $10^{\rm 7}$ yrars which is long enough for the
companion to become too dim to be detected in the supernova
remnant (\citealt{DISTEFANO12}). At the same time,
\citet{MENGXC17b} noticed that a spin-down timescale of
$\sim10^{\rm 6}$ years is favored to show the SD signal for the
SNe Ia from the CONe WDs.

In this paper, the method to calculate $\omega_{\rm f}$ is very
simple, and we do not solve the structure of the rapidly rotating
WD and follow the exact accretion history in detail. Then, the
exact $\omega_{\rm f}$ value of a WD may be different from that
shown in Figure~\ref{spin}. However, the trend of $\omega_{\rm f}$
with the initial WD mass will still hold, since our basic idea is
just from the conservation of angular momentum. In addition, we
assume that all SNe Ia explode at $M_{\rm WD}\sim1.4~M_{\odot}$,
while a rapidly rotating WD may exceed the mass. It has been
proved that, generally, the higher the initial mass of a WD, the
more massive the final mass is at the onset of the spin-down phase
and then the shorter the spin-down timescale (\citealt{WANGB14};
\citealt{HACHISU12}). Therefore, our assumption here could weaken
the effect of the initial mass on the spin-down timescale, but
does not change the trend. Moreover, the distribution of the
initial WD masses is based on the WD + MS channel, where the
distribution peaks at $M_{\rm WD}^{\rm i}\sim0.8~M_{\odot}$
(\citealt{MENGXC17a,MENGXC17b}), and the channels with red giant
(WD + RG) and helium star (WD + He star) companions are not
included in Figure \ref{mwd}. SN iPTF14atg is probably from the WD
+ RG channel and SN 2012Z is from WD + He star channel
(\citealt{CAOY15}; \citealt{MCCULLY14}). However, the distribution
of the initial WD masses from WD + RG channel is similar to that
from the WD + MS channel (\citealt{CHENXC11}). The peak of the
distribution from the WD + He star channel is higher than that
from WD + MS channel, i.e. at $M_{\rm WD}^{\rm i}\sim
1.0~M_{\odot}$, but the SNe Ia from the WD + He star channel may
only contribute to all SNe Ia by $\sim$10\%. Therefore, our
results may not be significantly affected by our simple
treatments, and our basic conclusion still holds. In particular,
the peculiar SN 2012Z also fulfills our suggestion, i.e. the
supernova is very likely derived from a massive hybrid CONe WD
from a binary
evolution point of view (\citealt{WANGB14b}). \\
\\
In summary, according to the spin-up/spin-down model, we propose a
universal explanation on why peculiar SNe Ia are more likely to
show the signature predicted by the SD model, while normal SNe Ia
are not. We suggest that this is derived from the different
initial masses and different initial chemical composition of the
WDs. Most of SNe Ia originate from the CO WDs, and these SNe Ia
experience a relatively long spin-down phase before supernova
explosion, which erases the SD signal. On the contrary, the
peculiar objects are suggested from the massive CONe WDs, and
these SNe Ia experience a shorter spin-down timescale than those
from the CO WDs before supernova explosion. A short spin-down
timescale means that these SNe Ia are more likely to show the
properties expected from the SD model. In particular, our model
suggests that some normal SNe Ia from massive CO WDs can show the
SD signal.

\section*{Acknowledgments}
This work was  supported by the NSFC (Nos. 11473063, 11522327,
11390374, 11521303, and 11733008), Yunnan Foundation (Nos.
2015HB096, 11733008), the CAS light of West China Program and CAS
(No. KJZD-EW-M06-01). Z.H. thanks the support by the Science and
Technology Innovation Talent Programme of the Yunnan Province (No.
2013HA005).

\end{document}